\documentclass[12pt]{iopart}

\usepackage{harvard}

\usepackage{graphicx}

\begin{document}

\title[MARCS models]{MARCS model atmospheres.}

\author{Bertrand Plez$^1,^2$}

\address{$^1$ GRAAL, CNRS, UMR5024, Universit\'e Montpellier 2, F-34095 Montpellier cedex 5, France}
\address{$^2$ Department of Physics and Astronomy, Uppsala University, SE-75120
Uppsala, Sweden}

\ead{bertrand.plez@graal.univ-montp2.fr}

\begin{abstract}
In this review presented at the Symposium {\sl A stellar journey} in Uppsala,
June 2008, I give my account of the historical  development of the MARCS
code from the first version published in 1975 and its premises 
to the 2008 grid. It is shown that the primary driver for the development
team is 
the science that can be done with the models, and that they constantly
strive to include the best possible physical data. A few preliminary 
comparisons of 
M star model spectra to spectrophotometric observations are presented.
Particular results related to opacity effects are discussed. The size of
errors in the spectral energy distribution (SED) and model thermal stratification are estimated for different  densities
of the wavelength sampling. The number of points used in the MARCS 2008 
grid (108000) is large enough to ensure errors of only a few K in all 
models of the grid, except the optically very thin layers of metal-poor stars.
Errors in SEDs may reach about 10\%\ locally in the UV. The published
sampled SEDs are thus appropriate to compute synthetic broad-band photometry, but 
higher resolution spectra will be computed in the near future and published 
as well on the MARCS site (marcs.astro.uu.se). Test model calculations
with TiO line opacity accounted for in scattering show an important
cooling of the upper atmospheric layers of red giants. Rough estimates
of radiative and collisional time scales for electronic transitions of TiO
indicate that scattering may well be the dominant mechanism in these lines.
However models constructed with this hypothesis are incompatible with 
optical observations of TiO (Arcturus) or IR observations of OH (Betelgeuse),
although they may succeed in explaining H$_2$O line observations. More 
work is needed in that direction.
\end{abstract}



\section{MARCS : little history}\label{history}
The MARCS model atmosphere code has been in use since the mid-70's. It
 has its roots in Bengt Gustafsson's early work
on a Feautrier-type method for model atmosphere calculation including convection 
\cite{1971A&A....10..187G}. A number of features, including detailed 
continuous opacities were already included in that pre-MARCS version. 
And in a Gustafsson way, it was directly applied to an astrophysical problem :
determining the metallicity of F type stars
 \cite{1972A&A....19..261G}.
A few years later the collaboration with Eriksson, Nordlund, and Bell 
gave birth to MARCS: a code for Model Atmospheres in Radiative and 
Convective Scheme \cite{1975A&A....42..407G,1976A&AS...23...37B}. 
It allowed the computation of hydrostatic, plane-parallel (PP),
line-blanketed atmospheres, with convection included following 
\possessivecite{1965ApJ...142..841H}
recipe for MLT. Line opacity was included in the form of Opacity Distribution Functions (ODF), and FGK type stars, also metal-poor were modelled. The 1975 paper is a highly recommended reading in the field,
as it discusses in great details the underlying assumptions, the algorithm, its implementation, and the resulting model structures with the impact of varying 
stellar parameters\footnote{incidentally, I was 12 then, and as far as I can remember 
this is about the time I got interested in astronomy, but the reason must 
be in a, at least initially, more contemplative approach to the stars...}.
The grid  was also used right away in astrophysical 
applications: abundance determinations, colour calibrations, ... 
This version of MARCS models have been used in almost 2 decades by the 
MARCS creators and collaborators, and probably longer by others.

Simultaneously, there were incentive to (i) update the code and its input
data, and (ii) extend its applicability to other spectral types.
The latter became effective with \possessivecite{1984A&A...132...37E}
efforts, who less than a decade later could deliver cool carbon star 
models with polyatomics included (HCN and C$_2$H$_2$), and apply them
to the determination of abundances of carbon stars \cite{1986ApJS...62..373L}. 
This development of MARCS was made possible because the expertise necessary to 
compute these new opacities was included into the team.
\citeasnoun{1984A&A...132...37E} could demonstrate the
strong effect of polyatomics opacities on the atmospheric structures and emergent
spectra, and already pointed out the necessity to refine these opacities 
\footnote{Thanks to the hard work of our colleagues in quantum chemistry this
problem is alleviated for HCN \cite{2002ApJ...578..657H,2006MNRAS.367..400H},
 but is unfortunately 
still pending today for C$_2$H$_2$. I again strongly 
encourage molecular physicists and quantum chemists to assemble a line
list for that species. There is a great reward : 
analysis of carbon AGB stars in the local group and the history of carbon, 
understanding of cool carbon dwarfs, ...}.
Two years later carbon star models were computed also with the Opacity Sampling
method by \citeasnoun{1986A&A...167..304E}. They demonstrated that the ODF 
hypothesis does not work well when opacities from different sources and not
correlated in wavelength (e.g. diatomics and polyatomics) dominate at various
atmospheric depths. The OS method is then more reliable provided a large 
enough number of sampling points are used (see \Sref{sampl}). At that time
computer limitations hampered the wide use of OS, but the trend was launched:
next MARCS would use OS.

The extension towards cool oxygen-rich stars is where it all started for me.
I had started my PhD in France on the empirical modelling of Mira stars based
on speckle interferometric observations. As I was getting about nowhere,
my supervisor had the good idea to send me to Uppsala, to spend about a year
with Bengt Gustafsson and learn as much as possible about cool star atmospheres.
The challenge became quickly to produce cool star atmospheres for oxygen-rich
stars, as only a carbon star setup was available\footnote{This turned out to demand
much more time, and I stayed in Uppsala a total of 3.5 years: first for 16 more
months as a  french civil servant part-time teaching French conversation to students and 
personnel of Uppsala university; then on a Swedish PhD fellowship. I met my 
wife during that time, and since then Sweden has become a second home-country to me.}. 
Opacities for TiO and H$_2$O
were missing, the chemical equilibrium did not include important species like
TiO, and sphericity was included in a trial version but did not work properly.
Thanks to Mats Larsson, Bosse Lindgren, and Lars Pettersson at Stockholm 
university I was introduced in the arcanes of quantum chemistry, and could 
produce a first line list for TiO. Together with John Brett we computed 
VO and CaH  line lists. We could not compute H$_2$O in the same manner, as
our limited quantum chemistry skills stopped at 2-atoms species, but used a 
trick to produce a pseudo line list from NASA observations of rocket exhausts
(mean opacities, and line density, together with a distribution of strengths).
We knew this was in need of improvement, as well as the other molecular opacity,
but it would allow, just as for carbon stars, a first assessment of the impact
of these species on the atmospheric structure.
The inclusion of spherical symmetry was made with the help of {\AA}ke Nordlund
using his algorithm \cite{1984mrt..book..211N}. Finally, The MARCS code was
made cool oxygen-rich capable by the modification of the chemical 
equilibrium routine, and OS was substituted to ODF with 11000 points, however keeping ODF as 
an option. The extensive Kurucz atomic line data was included by 
Bengt Edvardsson \citeaffixed{1993A&A...275..101E}{see}. In \citeasnoun{1992A&A...256..551P}, the 
improved code is described,
as well as the impact of the molecular opacities, and of sphericity on models. 
Low resolution spectra are compared to observations of M giants and the models
\cite{1992A&AS...94..527P}
are compared to other grids. Again these models were used with success 
in many applications in the following years.

Using the same code skeleton, \citeasnoun{1992A&A...261..263J} published
a grid of carbon stars, with 5400 OS points, sphericity, and C$_3$ opacity, although the latter 
would prove to be overestimated by a large factor later on.
They also discuss in details the effect of sphericity.

With the same code as \citeasnoun{1992A&AS...94..527P}, \citeasnoun{1993A&A...275..101E} 
computed a grid of metal-poor solar-type stars
with a combination of 4100 OS points in the UV ($\lambda<4500{\rm\AA}$) and
1400 ODF points in the red, and applied it to the chemical analysis of
Galactic disk stars, in a milestone paper.

The evolution continued towards more exotic spectral types with the
first grid of line-blanketed H-deficient models by \citeasnoun{1997A&A...318..521A} \citeaffixed{1997A&A...323..286A}{see also}.
Bound-free opacities were updated using data from the opacity project,
and free-free opacities were added for carbon and Helium ions. These were 
of course used to analyse RCrB and related stars \citeaffixed{1997A&A...321L..17A}{e.g.}.

There are a number of other papers summarising other add-ons and updates : 
\citeasnoun{1995A&A...295..736B} published a grid of M dwarfs, 
\citeasnoun{1998A&A...333..231B} \citeaffixed{1998A&A...337..321B}{see also}, 
in their work on the calibration of Johnson-Cousins photometry
used an updated MARCS grid of cool star models (both giants and dwarfs, including
updated molecular opacities, a much more extensive chemical equilibrium, 
and more OS points.
\citeasnoun{2003IAUS..210P..A2P} added ZrO opacities and computed the first
S-type star atmosphere grid with detailed blanketing (100000 OS points), and
made a first attempt at classifying S stars using the synthetic spectra and
 colours, \citeasnoun{2003ASPC..288..331G} published on the web 
 an extensive updated grid
of FGK type-stars down to very low metallicities. The two latter 
grids used a completely revised chemical equilibrium with 92 elements
and their ions as well as over 500 molecules, with updated partition functions and dissociation energies.

\section{MARCS 2008}\label{marcs2008}
The papers above, starting around 1998, were all paving the way, and announcing
the much awaited new MARCS\footnote{it was dubbed MARCS35 (guess why!) in 1998,
 and was supposed to be the final version for the new generation MARCS grid
while we were sketching the first draft of a series of papers, the first of which appeared only 10 years later...}. The idea of a large grid of updated 
MARCS models covering the cool part of the HR diagram dates back to a suggestion Bengt made to me in the park on Rackarberget at Dan Kiselman's PhD party in
 1993. Updates were constantly added, but culminated in the 2-3 years before 
 1998, while I was on the payroll of Uppsala Observatory
 again.
Citations of Gustafsson et al. in preparation date about back to 1998.
At some point we had a single version of the code, and a single set of physical input data,
but I got my job 
in Montpellier and the entropy started increasing again. This is without
mentioning the Copenhagen branch that  diverged already in 1991.
We had a number of small workshops or get together along the years to 
try to (i) get a single version of the code, (ii) update input data,
(iii) find tricks to avoid crashes of extreme models like high luminosity,
low gravity, cool supergiants, (iv) draft a series of papers and decide 
what to discuss in what paper. Then of course in between these meetings, other 
exciting projects, teaching and administration took over. 
During the past 9 months, my sabbatical in Uppsala allowed us once
more, and this time for good, to get a single more debugged, more updated (esp. 
continuous and line opacities) code from the Montpellier and Uppsala ones.
The first paper in a presumably long series appeared recently \cite{2008A&A...486..951G}
and the grid is (partly) on the web at marcs.astro.uu.se. 
MARCS 2008 is characterised by new opacities for H$_2$O, atomic collisional line broadening
included using the description of \citeasnoun{1995MNRAS.276..859A},
and hydrogen lines modelled
using a code by Barklem, described in \citeasnoun{2003IAUS..210P.E28B}. About
108000 OS points are used.
Full details are provided in the paper, that also relates some of the historical 
background, and discusses in depth the physical assumptions, numerical methods,
and physical data used. About 10000 models were computed \footnote{Thanks to this 
care for details and documentation that characterises Bengt's work, I could 
compare computing (CPU) times for the pre-MARCS code in 1972 : 25mn for a PP
model with 148 wavelengths accounting for the opacity of 25 Balmer lines, 
and MARCS 2008: 10mn for a spherical model with 108000 points accounting for 
over $10^8$ lines. The computing time is comparable, but the physics is 
much more detailed!}.

\section{Comparisons to observations}\label{comp}
\begin{figure}
   \centering
   \resizebox{\hsize}{!}{\includegraphics[angle=-90]{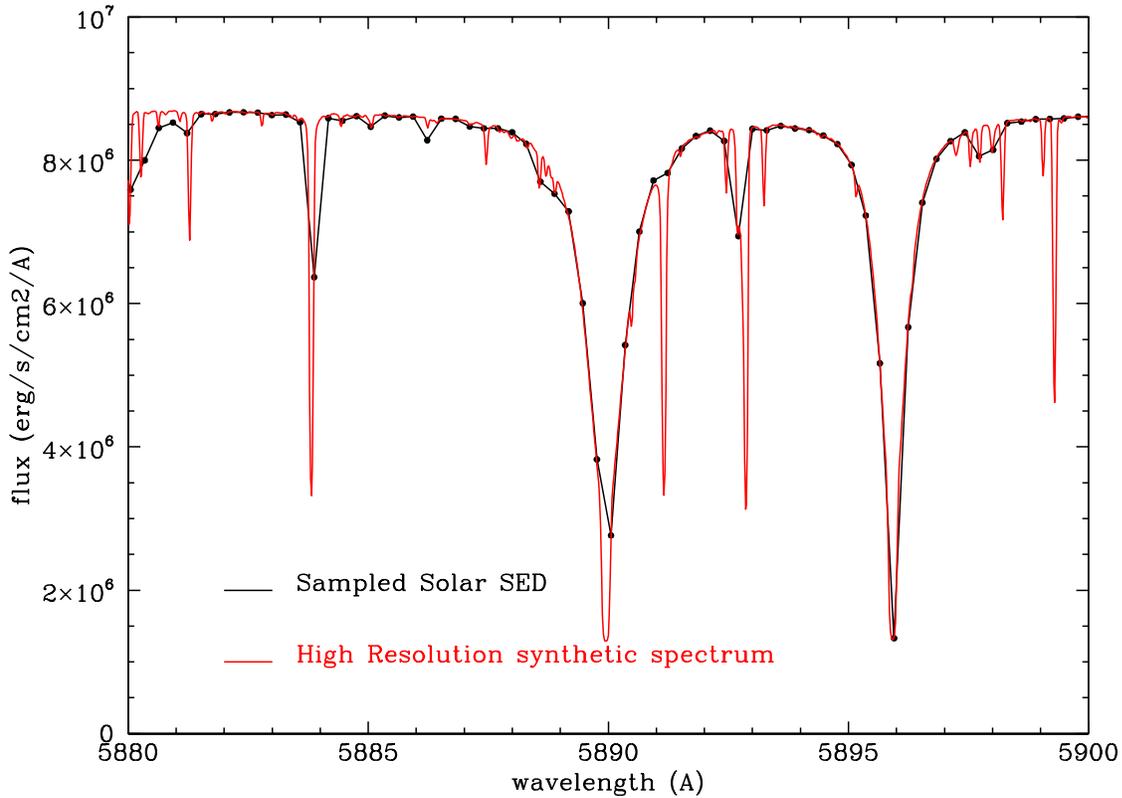}}
   \caption{Illustration of the fact that a sampled SED, as coming out from the
   MARCS code, is not equivalent to a spectrum at the same resolution. The 
   MARCS flux 
   is computed exactly at a number of prescribed wavelengths, but nothing is
   known about what happens between the points. Nevertheless when binned at
   sufficiently low resolution, the sampled SED tends towards the spectrum. }
   \label{OS-highres}
   \end{figure}

One decisive test to be passed by the models is a comparison to observed
high-resolution spectra and spectral energy distributions (SEDs). 
High-resolution spectra allow to check in detail if individual lines
are well accounted for in the model, whereas low-resolution SED 
comparisons permit to assess more easily problems in continuous opacity, 
thermal gradients, and missing opacities.
These comparisons must be carried out using  
stars with well determined stellar parameters
spanning the grid of models. In this volume, Bengt Edvardsson presents
the first comparisons to the Sun and solar-type stars SEDs. I have 
made a small number of comparisons for M-type dwarfs and giants.
The results I present here are very preliminary.
The model SEDs are the fluxes sampled at the 108000 OS points and as such
are not like an observed spectrum (see \Fref{OS-highres}). When averaging 
on a wavelength interval including many OS points, the mean average 
level of the spectrum is recovered as, statistically, the OS points hit 
the continuum and lines of all strengths. So, the sampled fluxes are 
a good representation of the SED only
when binned to a much lower resolution. Our OS scheme has a constant 
wavelength resolution of $\lambda/\Delta\lambda=20000$ between 900\AA\ and 20$\mu$m.
   
Comparisons were made with spectra from the MILES \cite{2006MNRAS.371..703S}
and STELIB \cite{2003A&A...402..433L} libraries, with both
observations and synthetic SEDs binned to 50\AA\ intervals. 
For the giants (M1 to M6III, i.e. 3800K to 3250K), the fit is generally
good, except for excess absorption in the models around 5100\AA\, most probably
due to MgH, and ups and downs shortward of 4200\AA\, reminiscent of what
Edvardsson (this volume) finds for the Sun (see \Sref{sampl} for a possible
explanation of part of this). At M6III, there is a hint of a missing opacity around 7500\AA\
(LaO?).
The dwarfs are more difficult. Using a $T_{\rm eff}$=3950K allows a fit
of the TiO bands of an M0.5V template, but the computed spectrum is then 
too blue. A lower temperature of 3750K gives a better shape for the continuum, 
at the expense of too deep TiO bands. In all cases there is excess blue flux
in the model. At M6V, TiO bands are fitted but there
 is also excess flux in the blue. Work with Mike Bessell in the few days
after this conference, using his own higher resolution spectrophotometry, 
allowed us to tie the blue flux problem to CaOH absorption,
and maybe AlH in metal-deficient stars, while providing better global fits. 
Detailed spectral comparisons need to be done, especially for dwarfs, missing
opacities should be assessed and if possible problem cured, and
a temperature scale based on the models  should  be derived, as has
been done for red supergiants \cite{2005ApJ...628..973L,2006ApJ...645.1102L}.
A paper
in the MARCS 2008 series will be devoted to M stars and will detail these
 comparisons. 

\section{Opacity effects}\label{opac}
A thorough discussion of the general effects of blanketing, of the
impact of different opacity sources, and of the effect of abundance and
microturbulence changes on blanketing in various 
temperature regimes of the MARCS models is presented in \citeasnoun{2008A&A...486..951G}. Here I will only complete this discussion
with two items: the impact of wavelength sampling, and some trial 
calculations on scattering in molecular lines.

\subsection{Wavelength sampling}\label{sampl}
I already stressed the fact that a large number of OS points is necessary 
in order to statistically well represent the opacity, i.e. to sample equally
well the continuum, and lines of all strengths. But what is {\sl large enough}?
There are two issues: (i) how many points are needed to give a good
representation of blanketing effects, and therefore provide a converged 
temperature stratification, in the sense that more OS points will not lead
to changes in the model stratification larger than a prescribed value (say 1K),
 (ii) how many points are needed to 
give a good representation of the SED, with a given model stratification
(\Fref{OS-highres}). 
These two numbers may well be different.
I computed series of models with subsets of our 108000 OS-points: 3 models 
with about 36000 points, 10 with 11000 points, and 30 with 3600 points. This 
was done for (i) a model of the Sun, (ii) a 3500K M giant, and (iii) a 5500K
metal-poor dwarf ([Fe/H]=-2). For each set of models, the average spectrum
and standard deviations were computed, and the same was done with the thermal
structure ($T-\tau_{\rm Ross}$).

In the case of the Sun, temperature fluctuations are less than 10K above
$\tau_{\rm Ross}$=-2.0, and less than 1 or 2K below, with 36000 OS points. They
are of the order of 30-50K above $\tau_{\rm Ross}$=-3.5, with 11000 points (up
to 100K for the most deviant model). Fluctuations stay below 15K in all cases
for $\tau_{\rm Ross}>-2.0$. The situation is much better for the cool giant,
with temperature errors less than 20K everywhere ($-6<\tau_{\rm Ross}<+2$, down to 3600 sampling points.
Errors decrease to less than 4K everywhere with 36000 points. 
On the contrary, the situation is worse in the optically thin regions 
for the metal-poor dwarf: temperature errors increase from 15K at 
$\tau_{\rm Ross}$=-4 to 120K at $\tau_{\rm Ross}$=-6, even with 36000 points.
The situation is better in the line forming region: less than 10K fluctuations
below $\tau_{\rm Ross}$=-2.0, at all samplings.
So, errors in the T-structure due to sampling, are only a few K in all cases, 
for sampling densities in excess of 36000 points. At smaller optical depths, 
errors may be large for metal-poor stars, and are small for cool giants ($<$10K). Errors in the temperature stratification occur only if the line
opacity is not well sampled, i.e. if the line and the OS-point densities 
are {\sl both} too low.

\begin{figure}
   \centering
   \resizebox{\hsize}{!}{\includegraphics[]{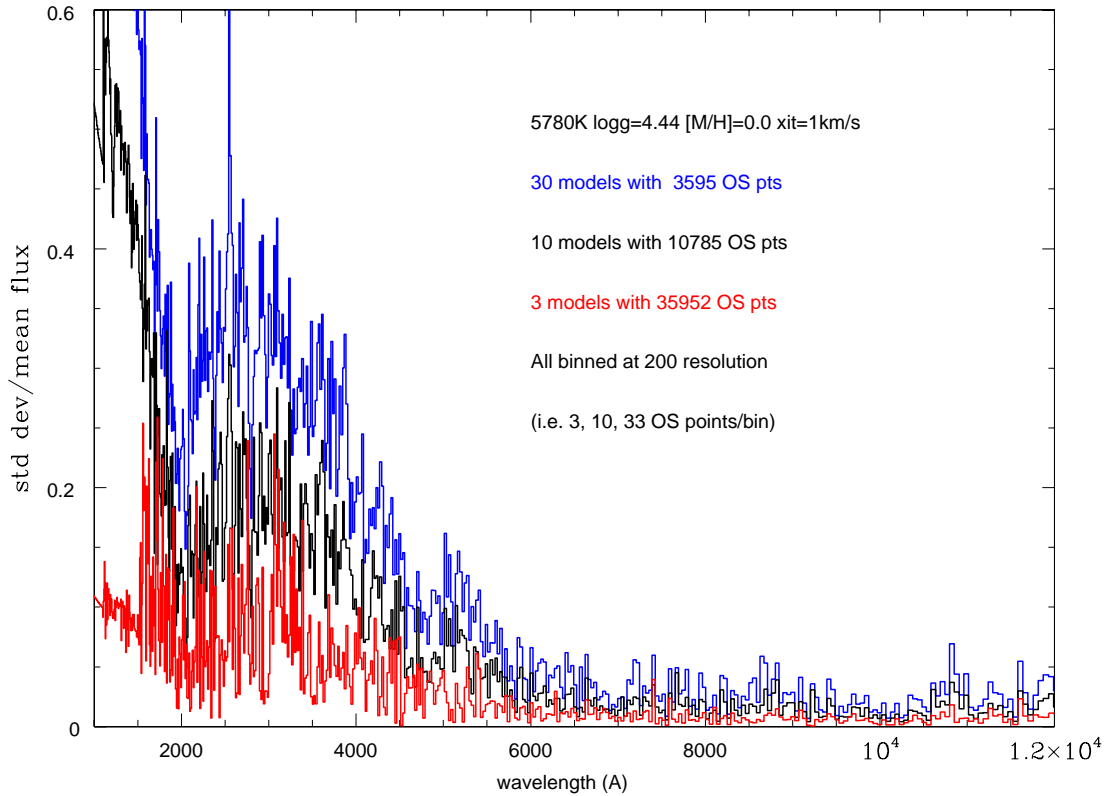}}
   \caption{Errors in sampled fluxes for Solar models computed with 
   different numbers of wavelength points. The figure shows the standard 
   deviation to the mean value of the flux in bins of 
   $\lambda\Delta\lambda=200$ for each series of models with 3600, 11000,
   or 36000 OS points.
   The reference MARCS 2008 models are computed with 108000 OS points.}
   \label{solarflux}
   \end{figure}

Errors in the fluxes are below a few percents everywhere the flux matters when 
using 36000 points, with
the notable exceptions of the UV flux of solar-type stars and the IR CO bands
of red giants. The case of the Sun is illustrated in \Fref{solarflux}. The 
errors around 3000\AA\ are of the order of 5 to 20\%, and around 5\% at 4000\AA.
The trend at all wavelengths when increasing from 3600 points to 36000 indicates
that the new MARCS grid with 108000 sampling points should provide sampled SEDs
with systematic errors $\sqrt 3$ smaller. The cool giant model sampled SED
shows errors of 5 to 10\% in the IR CO bands at 1.6 and 2.5$\mu$m, with 36000 OS-points. This is due
to the intense CO lines that are sparsely distributed, and sampled at 
a resolution of about 6500.
Not surprisingly, the flux of the metal-poor dwarf is very well modelled
even with a low sampling (there are almost no lines), except where there
are lines, i.e. in the UV. At the 36000 points sampling, errors reach 
more than 10\% only below about 1600\AA, and are less than 2 to 5\% between 
3000 and 4000\AA.

In conclusion, the 108000 OS points used in the computation of the MARCS 2008
models are sufficient for getting the errors on the thermal stratification
down to a few K for all model parameters of the grid. An exception is the optically very thin layers of metal-poor stars,
due to the fact that only a few lines dictate the thermal equilibrium
in these layers. This is however a region where NLTE or 3D effects 
most probably play a  greater role.
The situation is worse for the sampled SEDs that may still be off by over 10\%, 
esp. in the blue-UV, as many lines act on the spectrum without greatly
 affecting the temperature structure, e.g., atomic lines in cool giants, 
 see Fig. 5 in \cite{2008A&A...486..951G}, MgH or CH in the Sun. 
Of course better SEDs can be computed afterwards at very high resolution,
and this will be done for the MARCS models. Sampled SEDs can nevertheless 
be used, e.g., to compute broad band colours. 

\subsection{Scattering versus absorption}\label{scatt}
Molecular lines have a large effect on the thermal structure of cool stars,
through heating or cooling of the outer layers, and backwarming of deeper
layers. TiO numerous electronic transition lines in the optical are 
well known to lead to a large heating of the outer layers of cool stars.
This is due to this strong opacity appearing on the blue side of the local
source function and affecting thermal heating: 
$q=\int_0^{\infty}{\kappa_{\lambda}(J_{\lambda}-B_{\lambda})d\lambda}$, (with 
$q=0$ in LTE, hot radiation from below, $J_{\lambda}$, and the large 
$\kappa_{\lambda}$ in the blue impose an increase of $B_{\lambda}$
in the surface layers). Surface heating or cooling only happens if the opacity is 
in absorption, and a scattering term has no effect. It is well possible 
that some molecular lines do indeed form closer to scattering than pure 
absorption in the tenuous outer layers of red giants. \citeasnoun{1975MNRAS.170..447H} already suggested, based on meagre available 
collisional excitation data, that electronic transitions of molecules in cool
star atmospheres may be
dominated by radiative processes, and not by collisions with electrons or 
hydrogen. We may try to examine again the particular situation of TiO.
Radiative transitions occur with $t_{rad}=40~{\rm to}~100~{\rm ns}$ in most TiO 
electronic transitions. This means that collisional rates of excitation between
these levels must be at least $2.5 \times 10^7~{\rm s^{-1}}$ in order to compete and ensure LTE populations. For collisions with electrons if we use the, most probably
inappropriate but only available we have, approximations 
of \citeasnoun{1962ApJ...136..906V} and \citeasnoun{1968slf..book.....J} 
\citeaffixed{2003ASPC..288...99R}{see also}, an electron density of at least 
$3\times 10^{14} cm^{-3}$ is needed. In typical MARCS cool supergiant models
$2\times 10^6 < N_e < 2\times 10^{10}~{\rm cm}^{-3}$, well below the limit. For hydrogen,
\citeasnoun{1993PhST...47..186L} proposed a modification of 
\possessivecite{1969ZPhy..228...99D} formula, for 
hydrogen collisions with atoms that we may use, in lack of better
estimates, to derive a critical hydrogen density of about 
$2\times 10^{22}~{\rm m}^{-3}$. In the same MARCS models as above, we find
$2\times 10^{17} < N_H < 2\times 10^{21}~{\rm m}^{-3}$. Alternatively, using measured 
quenching rates in oxides, e.g. the measurements of \citeasnoun{Badieetal2008}
in YO, for Ar, He and O$_2$, we find a critical density of perturbers 
of about $10^{25}~{\rm m}^{-3}$ in order to reach collisional excitation 
rates of $2.5\times 10^7~{\rm s^{-1}}$.
The efficiency of  H collisions would have to be at least 5 orders of magnitude
larger than that of He for hydrogen to be of significance in the excitation
of YO electronic levels, and similarly of TiO. It seems thus very likely that radiative processes
indeed dominate over collisional processes in the population of electronic 
levels of TiO and in transitions between them. 

I therefore conducted a test calculation to study the impact on the thermal 
structure of red giant models if TiO lines were formed in scattering.
The first set of models had $T_{\rm eff}=4300~{\rm K}$, $\log {\rm g}=1.75$
[Fe/H]=-0.5, and M=1~M$_{\odot}$ representing Arcturus. The model with 
TiO in scattering has a temperature about 250~K lower at $\tau_{\rm 5000}=10^{-5}$, the difference decreasing to about 130~K at $\tau_{\rm 5000}=10^{-4}$ and vanishing
at $\tau_{\rm 5000}=10^{-3}$, the pressure stratification remaining unchanged.
This cooling is what \citeasnoun{2002ApJ...580..447R} propose to explain 
the appearance of H$_2$O lines at 12$\mu$m in Arcturus spectrum. However, 
when using the cooled MARCS model to generate a synthetic spectrum 
in the optical, TiO band heads (e.g. $\gamma (0,0)$ at 7054\AA) become
visible, whereas they remain undetected in \possessivecite{2000vnia.book.....H}
FTS spectrum of Arcturus. I did the same experiment with a model appropriate 
for Betelgeuse ($T_{\rm eff}=3600~{\rm K}$, $\log {\rm g}=0.0$,
[Fe/H]=0, and M=15~M$_{\odot}$. In this case the cooling amounts to about 250K
at  $\tau_{\rm 5000}=10^{-5}$, to about 130~K at $\tau_{\rm 5000}=10^{-3}$, 
and vanishes around $\tau_{\rm 5000}=2.5\times 10^{-2}$.
\citeasnoun{2006ApJ...637.1040R} detected the 12$\mu$m water vapour lines
in Betelgeuse as well but could not reproduce them with a standard 
atmosphere model at 3600~K, which is the temperature derived from the optical 
spectrum. They needed a much cooler (3250~K) atmosphere. Our cooled MARCS
model does not, however, reproduce the water lines : they remain too faint. In
addition neighbouring OH lines become too strong, but that could be alleviated by a decrease of O abundance. The 2.3$\mu$m CO bands
\cite{1996ApJS..107..312W}
are marginally better reproduced with the cooled MARCS model, but not better 
than using a molsphere, i.e. a spherical shell of gas at 2000~K, on top of
the photospheric model, as has been advocated by various authors to 
explain spectroscopic and high angular resolution observations of red 
supergiants in the IR 
\citeaffixed{2000ApJ...540L..99T,2007A&A...474..599P}{e.g.}.

Despite collisional time scale estimates being too long compared to radiative 
time scales to ensure LTE in 
electronic transitions of  TiO, the simple
assumption of its transitions occurring in scattering is not compatible 
with existing observations of red giants and supergiants. In particular 
it cannot solve the 12$\mu$m H$_2$O lines puzzle.
A better assessment of 
collisional rates in TiO, as well as a full NLTE treatment of the electronic
transitions, taking into account optical depth effects, as the lines may become 
optically very thick, would be of great value.
Also the coupling with hydrodynamics should be studied in detail as 
the balance between expansion cooling and radiative heating may be
shifted by large amounts in the upper atmospheric layers (see the
contribution by Wolfgang Hayek in this volume)

\section{Final thoughts}\label{conclusions}
After over 30 years, MARCS is now as mature as it can be within its approximations (hydrostatic, LTE, MLT). A much updated grid covering 
the red part of the HR diagram is now being published on the web 
(marcs.astro.uu.se), and the first paper in a series that will e.g. detail
the behaviour of all models, compare the synthetic spectra to observations
for a selected sample of template stars, or discuss synthetic photometry, 
has just appeared \cite{2008A&A...486..951G}. Although we are more and more 
turning 
our development activities towards relaxing some of the simplifying hypotheses
upon which MARCS rests, it is likely that updates of the MARCS grid will be made
available in the future, specially when better or additional  opacities can
be included. 
The drive behind the development of MARCS has always been the science that
can be done with it, and that is unlikely to change. It is a tool that 
despite eluding the dynamical side of reality, includes very detailed
accounts of opacities, and is flexible and light enough that it
can be used to test many ideas (effects of line scattering on atmospheric
stratification, impact of NLTE 
 over-ionisation on H$^-$ opacity, etc). In that respect it 
will remain irreplaceable for many years to come.

\ack
The MARCS code and its package of tools for e.g. detailed spectrum 
calculation is the result of many years (...decades!...) of 
collaborative, and mostly enjoyable work within the MARCS team: BEngt Edvardsson, Kjell Eriksson,
BenGt Gustafsson, {\AA}ke Nordlund, Uffe-Graae J{\o}rgensen, as well as a number of others whose names appear above. I am especially happy to thank BEngt, KjEll and BenGt for welcoming me during this year I spent on sabbatical in Uppsala.
It has been an exciting time, although of course I did not get done one third of
what I hoped to do, ... but the first MARCS 2008 paper is out!!
I thank Nils Ryde for discussions on molspheres and for helping me to 
get the best out of TEXES observations kindly provided by Mats Richter.
Most of all I wish to express my gratitude to BenGt, who in addition to
be a constant source of inspiration, showed me how to never be satisfied
with a result before I fully understand it. This may sometimes be very irritating, when projects or papers get stuck for months or years because of
such a difficulty, but the reward is greater in the end. Tack ! Och grattis
till f{\"o}delsedagen! 
\section*{References}
\bibliographystyle{jphysicsB}
\bibliography{plez}

\section*{Discussion}

\begin{description}
\item[Q:] (T. Lynas-Gray)
I note with interest the difficulty in matching energy distributions for 
M dwarfs. Did you use the water line list by Schwenke and Partridge, 
or the one by Barber and Tennyson?
\item[A:] The one by Barber and Tennyson.
\end{description}

\end{document}